\documentclass[traditabstract]{aa}
\usepackage{graphicx}
\usepackage{txfonts}
\usepackage{longtable}
\usepackage[authoryear]{natbib}
\bibpunct{(}{)}{;}{a}{}{,}

\def\ltsima{$\; \buildrel < \over \sim \;$}
\def\simlt{\lower.5ex\hbox{\ltsima}}
\def\gtsima{$\; \buildrel > \over \sim \;$}
\def\simgt{\lower.5ex\hbox{\gtsima}}
\newcommand{\HI}{\ion{H}{i} }
\newcommand{\HII}{\ion{H}{ii} }
\begin{document}
   \title{HST-ACS photometry of the isolated dwarf galaxy VV124=UGC4879}
\subtitle{Detection of the Blue Horizontal Branch and identification of two young star clusters\thanks{Based on observations made with the NASA/ESA Hubble Space Telescope, obtained at the Space Telescope Science Institute, which is operated by the Association of Universities for Research in Astronomy, Inc., under NASA contract NAS 5-26555. These observations are associated with program GO-11584 [P.I.: K. Chiboucas].}}

   \author{M. Bellazzini\inst{1}, S. Perina\inst{1}, S. Galleti\inst{1}, \and T. Oosterloo\inst{2}}
         
      \offprints{M. Bellazzini}

   \institute{INAF - Osservatorio Astronomico di Bologna,
              Via Ranzani 1, 40127 Bologna, Italy\\
            \email{michele.bellazzini@oabo.inaf.it} 
            \and
	    Netherlands Institute for Radio Astronomy (ASTRON), Postbus 2, 7990 AA Dwingeloo, The Netherlands\\
            Kapteyn Astronomical Institute, University of Groningen, Postbus 800, 9700 AV Groningen, The Netherlands
            }

     \authorrunning{M. Bellazzini et al.}
   \titlerunning{HST-ACS photometry of the dwarf galaxy VV124=UGC4879.}

   \date{Accepted for publication on A\&A }

\abstract{We present deep V and I  photometry of the isolated dwarf galaxy VV124=UGC4879,  obtained from archival images taken with the \emph{Hubble Space Telescope - Advanced Camera for 
Surveys}. 
In the color-magnitude diagrams of stars at distances larger than $40\arcsec$ from the 
center of the galaxy, we clearly identify for the first time a well-populated 
old Horizontal Branch (HB). We show that the distribution of these stars is more extended 
than that of Red Clump 
stars. This implies that very old and metal poor populations becomes more and 
more dominant in the outskirts of VV124. We also identify a 
massive ($M=1.2\pm 0.2\times10^4~M_{\sun}$) young (age$= 250\pm 50$~Myr) star 
cluster (C1), as well as another of younger age (C2, $\la 30\pm 10$~Myr) with a 
mass similar to classical open clusters  ($M\le 3.3\pm0.5\times10^3~M_{\sun}$). Both clusters lie at projected distances smaller than 100~pc from the center of the galaxy.\\}


\maketitle
%

\section{Introduction}
\label{intro}

The dwarf galaxy VV124=UGC4879 has been recently recognized to lie at the outer fringes of the Local Group (LG), near the turn-around radius \citep{k08,tik}. It is extremely isolated, being at $\simeq$1.3~Mpc from the nearest giant galaxy (the Milky Way) and at 
$\simeq$0.7~Mpc from the nearest dwarf (Leo~A). 

In \citet[][Pap-I hereafter]{vv124} we presented new deep wide-field photometry of VV124, which allowed us to confirm and refine the distance estimate by \citet{k08}, to study in some detail the stellar content of the galaxy and, in particular, its structure out to large radii, revealing the presence of low surface brightness structures in the outskirts. We also presented the first detection of \HI ~in VV124, obtained with the  Westerbork Synthesis Radio Telescope (WSRT). 
$10^6~M_{\sun}$ of \HI ~were found, with a maximum column density of $\ga 50\times10^{19}$~cm$^{-2}$.  
It is worth noting that the distribution of \HI ~is less extended than the stellar body of the galaxy.

\citet[][J11 hereafter]{j11} presented Hubble Space Telescope (HST) - Advanced Camera for Surveys (ACS) observations of VV124, resolving very well the stars down to $F814W\simeq 27$ also in the most central region of the galaxy. In particular, they studied the stellar content in the innermost $40\arcsec$ (see their Fig.~1), deriving the star formation history (SFH) in that region by means of the synthetic color-magnitude diamgram (CMD) technique \citep[see, for example][and references therein]{rizzi,cignoni}. J11 concluded that most of the SF ($\simeq$ 93  percent of the total, in mass) occurred at the earliest epochs, between 14 and 10 Gyr ago. After this, the SF ceased until 1 Gyr ago, when $\simeq$6 percent was formed between 1 and 0.5 Gyr ago and nearly 1 percent in the last 0.5 Gyr. The derived SFH is similar to that observed in other nearby dwarf irregular/transition-type galaxies \citep[see, for example][]{weisz}. The prevalence of old stars seems to increase at larger distances from the center (see also Pap-I).
From the morphology of the blue-loop region, J11 inferred that the metallicity of the youngest stars is $[M/H]=-0.7$.

In the present contribution, we use the same observational material of J11 to address three issues raised in Pap-I and not considered by J11. These are, in particular:

\begin{itemize}

\item The tentative detection of an extended Horizontal Branch (HB), including  stars in the color range of RR Lyrae variables and blue HB stars, in the CMD of the less-crowded outer regions of the galaxy (see Fig.~9 of Pap-I). While J11 report that a
population of HB stars reaching $V-I\sim 0.4$ is likely present in their CMD, its clean identification and quantification is hampered by the overlap with the young main sequence (MS) in the central region they consider. It would be interesting to search for the HB in the outer portion ($R_p>40\arcsec$) of the ACS field, where young stars should be quite rare and the old component should be dominant.

\item A possible correlation was noted between the asymmetric distributions of the young stars and of the neutral Hydrogen. Since the young population is concentrated in the innermost part of the galaxy, it can be traced in much deeper detail with the ACS data. This, in turn, prompts a more stringent comparison with the \HI distribution. 

\item We identified a candidate cluster located near the center of the galaxy and within the sheet of young stars (Cluster~1, C1 hereafter). The integrated spectrum of the source, shown in Pap-I, shows strong H$_{\beta}$ in absorption, suggesting a young age. With the high spatial resolution ACS image, we can hope to ascertain the real nature of C1 and look for other conspicuous clusters, if any. 

\end{itemize}

There is no reason to repeat the excellent analysis of J11, hence we focus strictly on the scientific cases listed above. 

\begin{figure}
   \centering
   \includegraphics[width=\columnwidth]{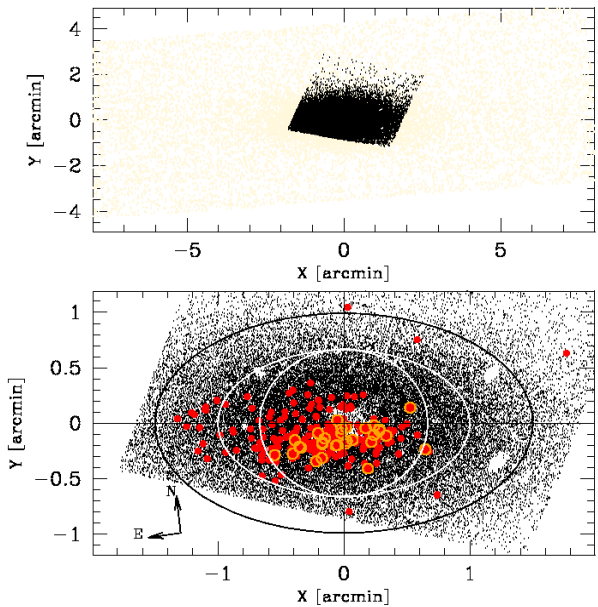}
     \caption{{\em Upper panel:} stars detected in the ACS images (dark dots) are superposed to the stars from the LBT sample of Pap-I (grey dots) in a projected coordinate system centered on the center of VV124 and rotated to have the X axis coinciding with the major axis of the galaxy (see Pap-I). Only stars brighter than $I=26.5$ from the ACS sample have been plotted, for clarity. {\em Lower panel:} zoomed view of the central part of the ACS sample in the same coordinate system. Two concentric ellipses of semi-major axis $a=1.0\arcmin$ and $a=1.5\arcmin$, having the same orientation and ellipticity of the galaxy isophotes (see Pap-I) are overplotted (in white and in black, respectively). The white circle has a radius $R_p=40\arcsec$, enclosing the region studied by J11. Blue stars lying on the young Main Sequence ($-0.4\le V-I\le 0.0$ and $I<25.5$, see Fig.~\ref{comcm}) are plotted as red filled circles, the brightest ones (having $I\le 23.0$) are encircled in orange.} 
        \label{map}
    \end{figure}

In the following, we adopt the parameters of VV124 derived in Pap-I (see their Tab.~1), if not stated otherwise. In particular, we adopt $D=1.3$~Mpc (in good agreement with J11) and $E(B-V)=0.015$ \citep[from][]{ebv}. At this distance $1\arcsec$ corresponds to 6.3~pc, $1\arcmin$ to 378~pc. We also adopt the de-projected coordinate system X,Y of Pap-I, centered on the center of VV124 and rotated such to let the X axis coincide with the major axis of the galaxy. Finally, we use the elliptical radial distance from the center of the galaxy $r_{\epsilon}$. To give an example, stars fulfilling the condition $r_{\epsilon}\le 1\arcmin$ are enclosed within an ellipse whose major axis coincides with the major axis of VV124 and has $a=1\arcmin$ and $\epsilon=0.44$ (see Fig.~\ref{map}, below). The normal (projected) radial distance from the center of the galaxy will be denoted with $R_p$.
All theoretical models used for comparison with observed CMDs are from the (canonical, solar-scaled) BASTI set \citep{basti,perci},  in the native ACS- Wide Field Channel (WFC) VEGAMAG 
system\footnote{\tt http://193.204.1.62/index.html}.

   \begin{figure}
   \centering
   \includegraphics[width=\columnwidth]{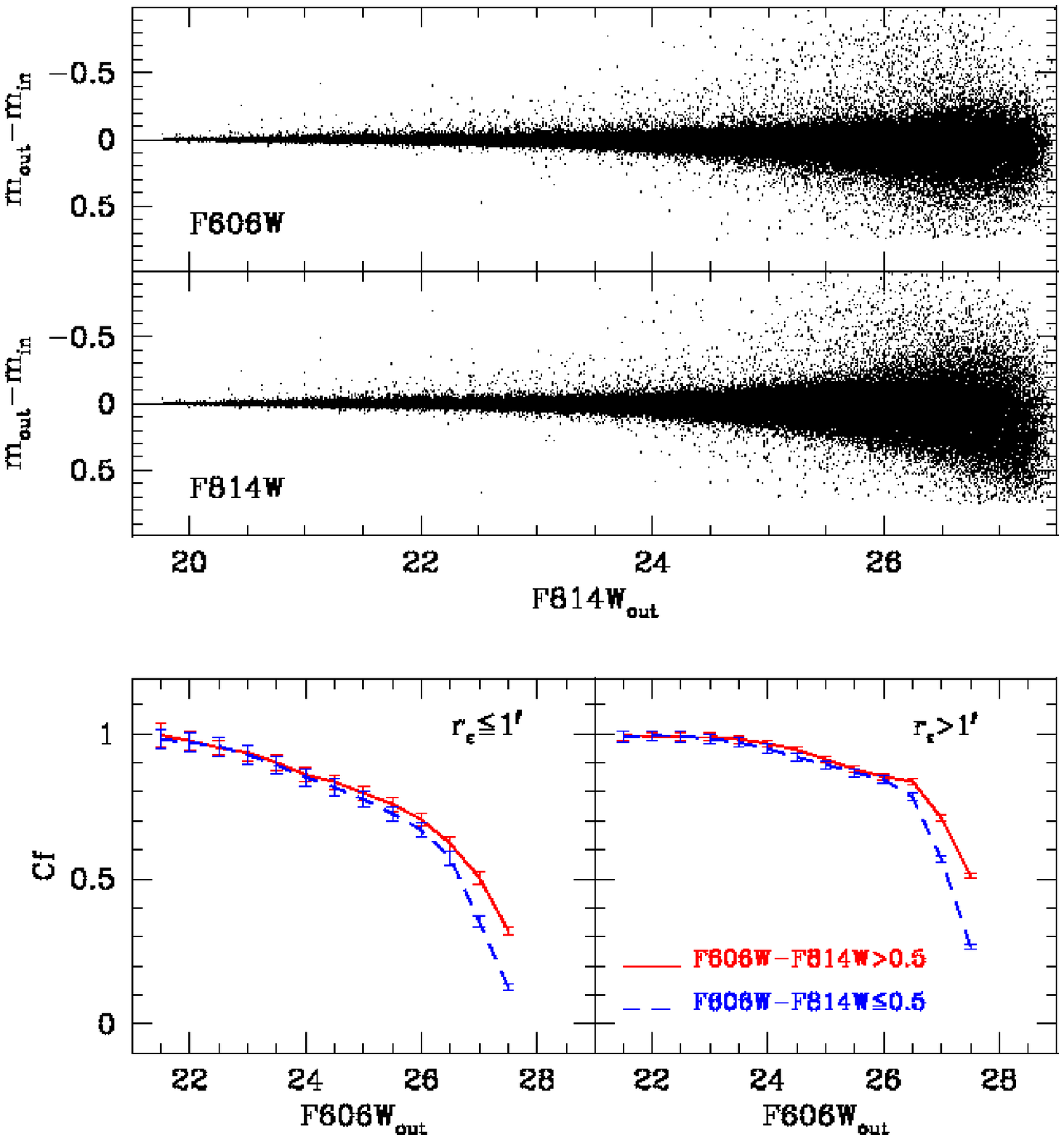}
     \caption{Results of the artificial stars experiments. {\em Upper panels: } Difference between the {\em output} and {\em input} magnitudes of the artificial stars as a function of F814W magnitude. {\em Lower panels:} Completeness as a function of F606W magnitude in the inner ($r\le 1.0\arcmin$; left panel) and outer ($r> 1.0\arcmin$; right panel) regions of the galaxy. Red continuous curves shows the completeness for stars in the color range of the RGB and the RC, blue dashed curves refer to stars in the color range of the old HB and the YMS (see Fig.~\ref{comcm}, below). The blue curves start at $F606W=25.0$ because in that color range we produced only artificial stars fainter than this limit, focussing to the study of the completeness in the blue HB region.} 
        \label{CFplot}
    \end{figure}


The plan of the paper is as follows: in Sect.~\ref{obs} we describe the observational material and the data reduction process; in Sect.~\ref{bhb} we present the clear detection of the extended HB in the external regions of the ACS field and we discuss the radial distribution of the various stellar populations of VV124. Sect.~\ref{sec:HI} is devoted to a brief discussion of the correlation between the spatial distribution of the \HI and of the young stars, and in Sect.~\ref{clus} we study the candidate luminous star clusters identified here and in Pap-I. Finally we briefly summarize and discuss our main results in Sect.~\ref{conc}.

\section{Observations and data reduction}
\label{obs}

The observational material consists of two pairs of F606W and F814W ACS-WFC\footnote{\tt http://www.stsci.edu/hst/acs/} images  obtained within the program GO-11584, which we retrieved from the HST archive. The total exposure time of the stacked drizzled images is 1140~s and 1150~s for the F606W and F814W images, respectively. The WFC is a mosaic of two $4096 \times 2048$ px$^2$ CCDs, with a pixel scale of 0.05 arcsec~px$^{-1}$. The center of VV124 is placed at the center of one of the mosaic chips (chip~1).

Positions and photometry of individual stars were obtained with the ACS module of the Point Spread Function (PSF) - fitting package DOLPHOT v.1.1\footnote{\tt \tiny http://purcell.as.arizona.edu/dolphot/} \citep{hstphot}, as described, for example, in 
\citet{b514}. The package identifies the sources on a reference image \citep[in this case the drizzled F814W image, as in][]{mack} and performs the photometry on individual frames, also taking into account all the information about image cosmetics and cosmic-ray hits which is attached to the observational material. 
We adopted a threshold of $3\sigma$ above the background for the source detection. 
DOLPHOT automatically provides magnitudes both in the WFC VEGAMAG system
and in the Johnson-Kron-Cousins system \citep{hstphot}.
As done in \citet{b514} and \citet{vdb0,all}, we selected the sources according to the quality parameters provided by DOLPHOT, to clean the sample from spurious or badly measured entries. In particular, we retained in the final catalogue only stars having {\em quality flag}=1 (i.e.\ best-measured stars), ${\tt chi}<2.0$, $|{\tt sharp}|<0.5$,  crowding parameter ${\tt crowd}<0.3$\footnote{We will relax this selection criterion to ${\tt crowd}<1.0$ in Sect.~5, to study in more detail the star clusters, where stars are significantly more crowded than average.}, and photometric error $<0.3$~mag in both filters \citep[see][for details on the meaning of the various parameters]{hstphot}. 
In the following sections we will always adopt this selected catalogue that contains 44894 sources, in total \footnote{This catalogue is available in electronic form at the CDS 
via anonymous ftp to cdsarc.u-strasbg.fr (130.79.128.5) or 
via http://cdsweb.u-strasbg.fr/cgi-bin/qcat?J/A+A/.../...}.

   \begin{figure*}
   \centering
   \includegraphics[width=\textwidth]{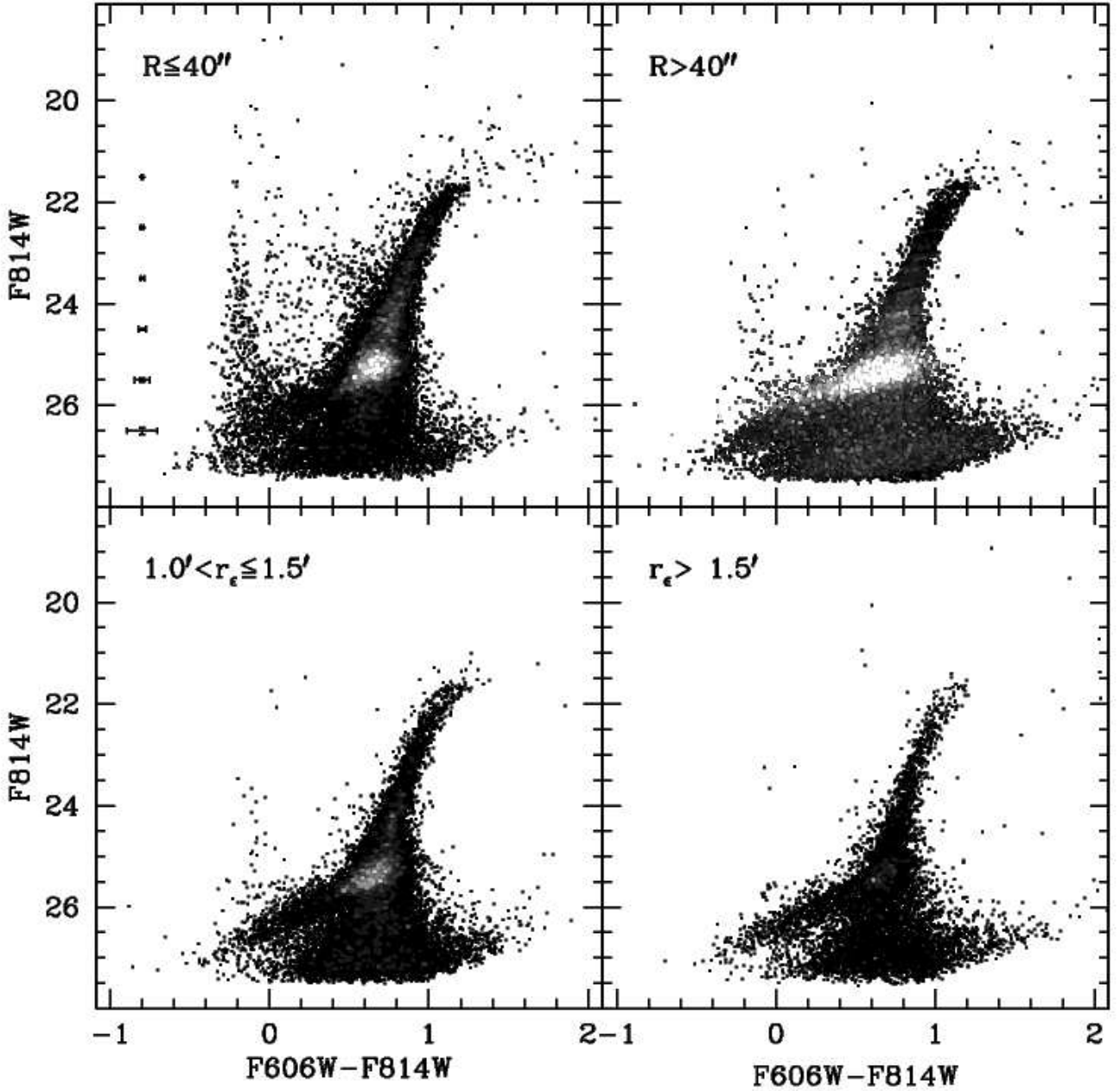}
     \caption{{\em Upper panels:} CMD of VV124 from the ACS data within (left panel) and outside (right panel) a circle of radius $R_p=40\arcsec$. The average photometric errors are reported as error bars in the left panel. {\em Lower panels:} CMDs in other regions of the ACS field delimited by the ellipses plotted in Fig.~\ref{map}. The right panel regards a radial region similar to that considered in Pap-I to tentatively detect the extended HB population (see their Fig.~9). In all the panels stars are plotted as black points in regions of the CMD with few stars and as grey squares otherwise, with the scale of grey proportional to the local density of stars in the CMD (lighter tones of grey correspond to higher density).}
        \label{comcm}
    \end{figure*}


We transformed the star positions from this system into J2000 RA and Dec by fitting a fourth degree polynomial to 5038 stars in common between the ACS catalogue and the LBT catalogue of Pap-I, using the CataPack suite of codes\footnote{Developed by Paolo Montegriffo at INAF - Osservatorio Astronomico di Bologna.}. The r.m.s.\ of the residual was $0.07\arcsec$ in both RA and Dec. The positions of the stars in the ACS field with respect to the LBT sample are shown in the upper panel of Fig.~\ref{map}. In the lower panel of the same figure we zoom on the part of the ACS field containing the center of the galaxy. It is particularly interesting to note that a large number of stars are detected beyond the $R_p\le 40\arcsec$ limit of the sample used by J11 for their determination of the SFH. It is beyond that radius, and especially beyond $r_{\epsilon}=1.5\arcmin$ where we expect to detect the extended HB of VV124, an obvious tracer of very old (age$\ga 10$ Gyr) and metal-poor populations. 

\subsection{Artificial stars experiments}

 To study the completeness of the sample and the photometric uncertainties, we performed a set of artificial star experiments, following the same procedure adopted by \citet[][see that paper for a detailed description and references]{vdb0}. We produced $\simeq 200000$ artificial stars having $21.0\le F606W\le 28.0$, with colors in the range covered by the bulk of observed stars, i.e.\ $-0.4\le F606W-F814W\le 1.3$. 

The results of the experiments are presented in Fig.~\ref{CFplot}. The upper panels show that the photometric accuracy is better than 0.1 mag for $F606W\la 25.0$; a quantitative summary of the results displayed in these panels is presented in Table~\ref{Tab_errart}.
The lower panels show the completeness factor ($C_f$) as a function of magnitude for two radial ranges and for two color ranges.
At magnitudes fainter than $F606W=25.0$, the completeness is slightly worse for blue stars than for red stars. However, even blue stars in the inner radial bin have $C_f\ga$50\% for $F606W\la26.5$. In the outer radial bin, the same level of completeness is reached at $F606W\simeq 27.2$.

\begin{table}
  \begin{center}
  \caption{Photometric errors from artificial stars experiments}\label{Tab_errart}
  \begin{tabular}{lcc}
    \hline
    F814W & $\sigma(F814W_o-F814W_i)$ & $\sigma(col_o-col_i)^a$\\
\hline
20.0$^b$ & 0.009	    & 0.006	   \\
21.0 & 0.010	    & 0.009	   \\
22.0 & 0.013	    & 0.014	   \\
23.0 & 0.021	    & 0.024	   \\
24.0 & 0.040(0.045)$^c$ & 0.044(0.045) \\
25.0 & 0.077(0.085) & 0.084(0.087) \\
26.0 & 0.153(0.164) & 0.175(0.180) \\
27.0 & 0.236(0.242) & 0.281(0.286) \\
\hline
\end{tabular}
\tablefoot{
$^a$ $col_o=F606W_o-F814W_o$; $col_i=F606W_i-F814W_i$.\\
$^b$ The reported errors are standard deviations computed for recovered artificial stars having $F814W_o$
within $\pm 0.5$ mag from the value reported in column 1.\\
$^c$ Numbers within parentheses refer to the sample with $CRO\le 1.0$, which is adopted
in the analysis of the clusters. All the other values refer to the standard $CRO\le 0.3$ sample.
The alternative value is reported only in the cases when it differs by more than 0.001 mag from the standard value.
} 
\end{center}
\end{table}

\section{The extended HB of VV124}
\label{bhb}

In the upper left panel of Fig.~\ref{comcm} we show the ACS CMD of VV124 in the same radial range studied by J11 ($R_p\le 40\arcsec$). The two CMDs are very similar, as expected, since they are obtained from the same dataset and with the same photometry package. 

The most obvious features are the broad RGB, with the RGB tip located at $F814W\sim 21.5$, and the Red Clump of intermediate/low mass core He burning stars around $F814W\sim 25.3$. A handful of Asymptotic Giant Branch stars is present on the extension of the RGB above the tip.
The obvious blue plume of young MS stars runs  over all the considered magnitude range around $F606W-F814W\sim -0.2$. The sparse plume protruding from the RGB at $F814W\sim 23.5$ toward the blue has been identified by J11 as composed of Blue Loop stars of metallicity [Fe/H]$\sim -0.7$; these are the evolved counterpart of the young MS. 

J11 noted that the synthetic CMD best reproducing the observations included HB stars reaching colors as blue as $V-I\sim F606W-F814W\simeq 0.4$ that were not identified in the observed CMD. We note, on the other hand, that it seems possible to discern a blue extension of the HB  in our CMD around $F814W\sim 25.8$ and $0.0\la F606W-F814W\la 0.6$. This impression is confirmed by the fact that the CMD of stars beyond the region studied by J11 display an obvious extended HB (upper right panel of Fig.~\ref{comcm}), with an higher-density portion between $F606W-F814W\simeq 0.6$ and $F606W-F814W\simeq0.3$, and a less populated part reaching $F606W-F814W\simeq -0.4$.

The extended HB becomes more evident when more external regions are considered, as shown in the lower panels of Fig.~\ref{comcm}. This should be partially due to the decreased crowding, which allows to get more accurate photometry, thus decreasing the width of the other CMD sequences that may partially overlap with the HB. However, the most important effect is the decreasing impact of the young MS plume, which contributes to hide the blue HB near the center of VV124. We will show below that, in addition to these observational factors, there is also a genuine population gradient making the HB stars more abundant in the outer regions of the galaxy.
The lower right panel of the figure shows the CMD in the same radial region where we obtained the tentative detection of the extended HB from the LBT data, thus providing a direct confirmation of that finding.

   \begin{figure}
   \centering
   \includegraphics[width=\columnwidth]{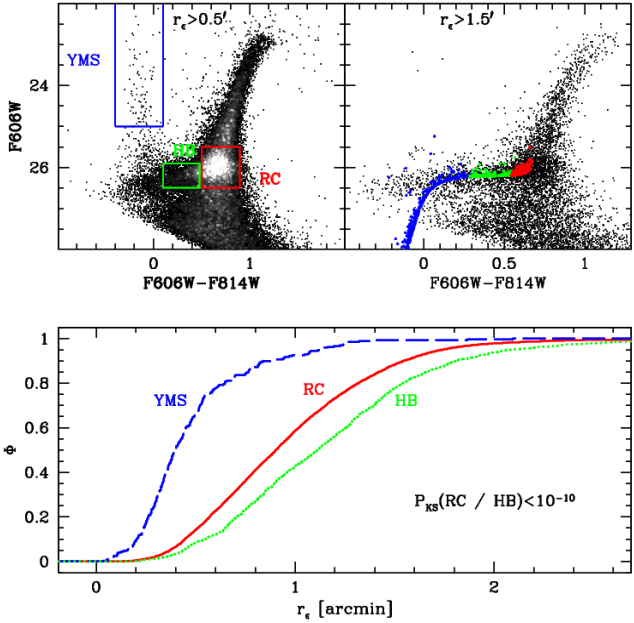}
     \caption{{\em Upper left panel:} The adopted selection boxes are plotted on the CMD for $r_{\epsilon}>0.5\arcmin$ (the innermost region has been excluded to avoid confusion in the panel). {\em Upper right panel:} A synthetic HB population computed from BASTI models, reported to the same distance and reddening of VV124, is superposed to the observed CMD for $r_{\epsilon}>1.5\arcmin$. The radial selection is adopted to obtain the cleanest view of the HB. Red HB, RR Lyrae variables, and Blue HB are plotted in red, green and blue, respectively. {\em Lower panel:} Radial distribution of the various populations selected with the CMD boxes, as indicated by the labels.  } 
        \label{crad}
    \end{figure}


In the upper right panel of Fig.~\ref{crad}, we superposed a synthetic HB population, produced from the BASTI theoretical models \citep{basti} with the dedicated web tool\footnote{\tt http://193.204.1.62/index.html}, to the observed CMD of the outermost region of VV124 sampled by the ACS data. The synthetic population (1000 stars) have $Z=0.001$ \citep[appropriate for the old population of the galaxy, according to][Pap-I and J11]{k08}, a mean mass of 0.494~$M_{\sun}$ and a dispersion $\sigma=0.2~M_{\sun}$, and it has been shifted to the distance and reddening of VV124. This is not intended to reproduce the observed HB but only to provide a sanity check for the interpretation of the sequence and to have a hint on the classification of stars within the HB (color coded according to the classification provided by the model). In particular, it is evident that the galaxy likely host a significant population of RR Lyrae variables, since there are many HB stars in the color range typical of these variables. Clearly their  definitive classification as genuine RR Lyrae would need time series photometry, which is not available to us.
The detection of such a rich population of candidate RR Lyrae and blue HB stars is fully compatible with the SFH derived by J11, who concluded that $\ga$~90\% of the stars were formed between 14 and 10 Gyr ago.

\begin{figure}
   \centering
   \includegraphics[width=\columnwidth]{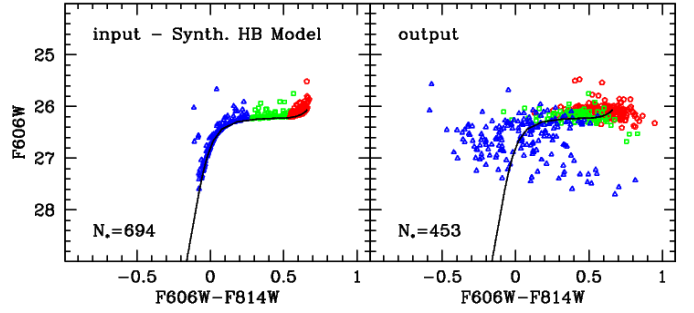}
     \caption{ Left panel: CMD of the synthetic stars from the BASTI model shown in Fig.~\ref{crad} having $25.5\le F606W\le28.0$. The meaning of the symbols is the same as in Fig.~\ref{crad}. The continuous line is the corresponding Zero Age HB model, plotted in both panels as a reference. Right panel: simulation of the observation process on the synthetic stars, obtained from the artificial stars experiments (sample with $r_{\epsilon}>1.5\arcmin$). Note that the output HB morphology is very similar to the observed one.} 
        \label{ossHB}
    \end{figure}

For F606W-F814W$\la 0.0$, the bluest portion of the observed HB seems to depart from the behavior predicted by the models, showing only a modest decline of the mean magnitude with color, while the synthetic population plunges towards much fainter luminosity. This may be due to a combination of effects, including non-perfect temperature-color transformations in this extreme color regime and photometric errors/completeness/blending/selection effects due to the proximity of the limiting magnitude level, which becomes brighter for bluer colors. In Fig.~\ref{ossHB} we simulate the observational effects on a subsample of the synthetic HB stars. The experiment demonstrates that the apparent mismatch can be fully accounted for by the effects of photometric errors and blending. Hence, most of the stars in the horizontal strip around $F606W\sim 26.5$ and bluer than $F606W-F814W\sim 0.0$ are genuine BHB stars (at least for $r_{\epsilon}>1.5\arcmin$, where the contamination by YMS stars is negligible), in spite of the apparent mismatch with the raw models\footnote{To simulate the observations we proceeded as follows. For each model star plotted in the left panel of Fig.~\ref{ossHB}, we isolated the subset of artificial stars having input F606W magnitude within $\pm 0.2$ and input color within $\pm 0.05$ of the considered synthetic star. Then we extracted at random one of the selected artificial stars. If the extracted artificial star is not recovered in the output we remove the corresponding synthetic star, thus mimicking the effect of incompleteness. If it is recovered, we compute its output-input difference in magnitude and color and we sum these quantities to the magnitude and color of the synthetic star, thus mimicking the effects of photometric errors and blending \citep[see][for a similar application]{n288}. We repeated the experiment several times and we verified that the HB morphology in output does not depend on the specific random realization. For an example of detailed modeling of an HB population in a similar context see \citet[][and references therein]{monelli}.}.


The upper left panel of Fig.~\ref{crad} shows the boxes we adopted to select stars on the Young Main Sequence (YMS), in the Red Clump (RC) and in the extended HB, to compare their spatial distributions. For the HB we adopted a box covering a relatively narrow color range, approximately matching the one of RR Lyrae variables. This choice was forced by the need to avoid contamination from RC stars, on the red side, and, above all, from faint YMS stars on the blue side. This would have been particularly problematic since (a) YMS are obviously much more centrally concentrated than older stars, and (b) the YMS blue plume becomes wider and reaches redder colors at faint magnitudes ($V\la 25$), clearly overlapping with the blue HB in the central region. We verified that the actual radial distribution of HB stars bluer than the adopted selection box is fully compatible with a population having the same radial distribution as the HB sample plus a strong contamination of YMS in the innermost $\sim 0.5\arcmin$. Hence the selected HB sample can be considered as a fair tracer of the old HB population as a whole, including BHB.
The selected RC and HB samples should have similar completeness properties, since they have similar magnitudes and colors. To verify this hypothesis we 
compared the behavior of the completeness as a function of $r_{\epsilon}$ for the two samples, using our set of artificial stars. It turned out that the two completeness curves are undistinguishable: hence any observed difference in the radial distribution of RC and HB stars cannot be accounted for by completeness effects.

The cumulative radial distributions of the three selected populations are compared in the lower panel of Fig.~\ref{crad}. In addition to the striking (and already noted) difference between the YMS and the older RC and HB populations, it is clearly revealed that the HB stars have a more extended distribution with respect to RC stars. A Kolmogorov-Smirnov test gives that the probability that the two samples are extracted from the same parent population is $P<\times10^{-10}$, thus firmly establishing the significance of the difference. It must be concluded that VV124 displays the classical population gradient observed in all dwarf galaxies, with the older/more-metal-poor populations becoming increasingly more abundant at larger radii and younger/more-metal-rich being more concentrated toward the center \citep{harbeck,umi,leo2,tht}.

\subsection{The color distribution of RGB stars}

\begin{figure}
   \centering
   \includegraphics[width=\columnwidth]{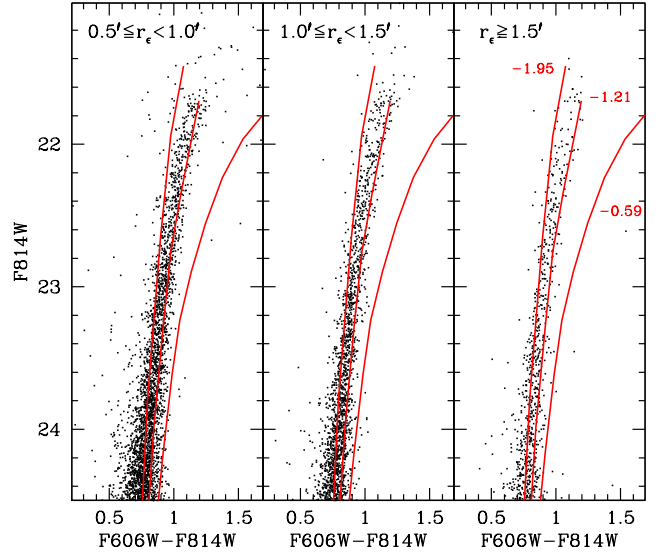}
     \caption{Comparison of the CMDs of VV124 in different (elliptical) radial annuli with the RGB ridge lines for the Galactic globular clusters NGC 6341, NGC 6752 and NGC 104 (from left to right), from the set by \citet[][]{bedin}, that provides the templates in the same ACS-WFC VEGAMAG system as our observations. The reddening, distance moduli and metallicities of the clusters are taken from \citet{f99}, as in Pap-I. In the right panel the ridge lines are labeled with the [M/H] value of the corresponding cluster in the \citet{cg97} scale \citep[col.~4 of Tab.~2 of][]{f99}. The radial range of the left panel coincides with the innermost radial range considered in Pap-I.} 
        \label{cmlm}
    \end{figure}

In Pap-I, the comparison of the observed RGB with empirical templates from Galactic globular clusters revealed the possible presence of a small metallicity gradient in the same sense.  The analogous comparison, performed with the much more accurate ACS photometry is presented in Fig.~\ref{cmlm}, to
check the findings of Pap-I and to verify whether the distribution and mean color of RGB stars is compatible with the gradient found with the HB stars. It must be stressed that the following analysis is not intended to provide accurate metallicity estimates, since it is well known that photometric determinations of the metallicity can be affected by serious degeneracies when the underlying age distribution is not well constrained \citep[see][and references therein]{cole,lianou}. On the other hand we look for differences in the inner and outer samples of RGB stars that may or may not be in agreement with the observed HB gradient. In the analysis we will adopt the {\em global metallicity parameter} [M/H] introduced by \citet{scs} to allow meaningful comparisons between the RGB colors of stellar populations having different $\alpha$-element abundances \citep[see also][]{f99,umi,santi}.

Fig.~\ref{cmlm} shows that the fraction of stars spilling to the red of the [M/H]=-1.21 template is indeed larger in the $0.5\arcmin\le r_{\epsilon}<1.0\arcmin$ range than at larger distances, even if the amplitude of the color spread is smaller than that apparent in Pap-I. Considering only stars in the range $21.7\le F814W\le 23.0$ (i.e.\ the least affected by photometric errors and crowding effects) and enclosed between the bluest and the reddest template, we computed the fraction of stars lying to the red of the [M/H]=-1.21 ridge line. This is $0.29\pm 0.03$ in the range $0.5\arcmin\le r_{\epsilon}<1.0\arcmin$ and $0.20\pm 0.02$ for $r_{\epsilon}\ge 1.0\arcmin$ (the same result is obtained also if stars with $r_{\epsilon}<0.5\arcmin$ are included in the inner bin, and if stars down to $F814W=23.5$ are considered).  This is quite similar to the results obtained in Pap-I and in broad agreement with the gradient shown in 
Fig.~\ref{crad}.

We obtained metallicity estimates for individual stars with F814W$\le 23.0$, by linear interpolation within the templates grid, as in Pap-I. The resulting metallicity distributions are quite broad, ranging from the blue limit of the grid, [M/H]=-1.95, to [M/H]$\sim -1.0$. The mean metallicity in the range $0.5\arcmin\le r_{\epsilon}<1.0\arcmin$ is $\langle {\rm[M/H]}\rangle =-1.4$, while for $r_{\epsilon}\ge 1.0\arcmin$ it is $\langle {\rm[M/H]}\rangle =-1.5$.
 The associated standard deviation, deconvolved from the observational effects, is $\sigma_{\rm[M/H]}\simeq 0.25$~dex in both radial bins, also in reasonable agreement with the results of Pap-I.


\section{The spatial distributions of \HI and young stars}
\label{sec:HI}

\begin{figure}
   \centering
   \includegraphics[width=\columnwidth]{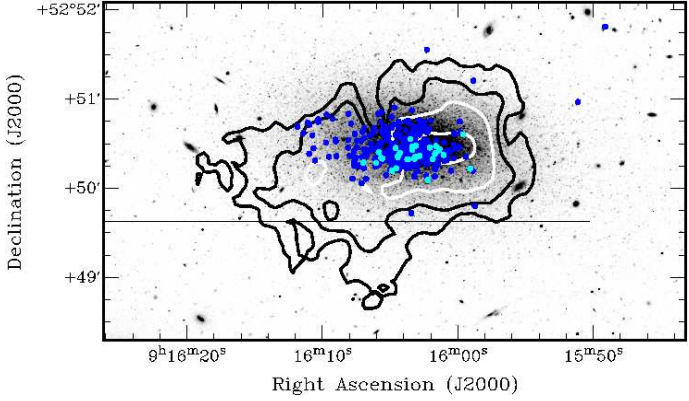}
     \caption{Contours of the total HI image drawn on
top of an optical $g$ band image obtained with the LBT, both from Pap-I. Contour levels are 5, 10 (black contours), 20 and 50 (white
contours) $\times 10^{19}$ cm$^{-2}$ . Blue circles denote the stars with
$-0.4 < V-I < 0.0$ and $23 < I < 25.5$ while the cyan circles show the positions
of the stars in the same color interval but with $ I < 23$, as if
Fig.~\ref{map}).} 
        \label{HI}
    \end{figure}

In Figure \ref{HI} we compare the location of those stars identified to be on
the YMS with the \HI\ distribution. This comparison clearly shows that,  when
considered on large scales, there is a very good spatial correspondence between the
distribution of the YMS stars and that of the \HI, both show the same asymmetry
with respect to the overall light distribution of VV124.  Such an overall
correspondence is not entirely unexpected, because, after all, stars form from
gas.  On smaller  scales (i.e.\ less than 100 pc) the spatial correspondence
between YMS stars and the \HI\ is less good, with the peak of the \HI\ somewhat
displaced from the YMS stars. This situation is not untypical for small
galaxies:  a good correlation between star formation and \HI\ on large scales
and the breakdown of it on smaller scales is observed in many small galaxies
\citep{Begum06,Roy09}.

The close overall correspondence between the YMS and the \HI\ also suggests that
the observed kinematics of the \HI\ is also connected to the formation of the
stars we now observe to be on the YMS. As discussed in Pap-I, if the
\HI\ tail is interpreted as an outflow, the energy output of roughly a dozen OB
stars would be sufficient to drive it. However, an interesting alternative is
that the tail corresponds to infall of gas and that VV124 is re-accreting part
of the gas which was expelled by the formation of the stars we now see as an old
population. That such re-accretion may occur is suggested by some recent simulations \citep[e.g.][]{S11}, and would also be very well consistent with the  fact that VV124  has been dormant for a long period of time since the early phase of star formation and only very recently started to form young stars again. The metallicity of the young stars in VV124 ([Fe/H]$\sim -0.7$, as estimated by J11 from the observed morphology of the Blue Loop) also suggest that, if they formed from the HI gas cloud, this cloud must contain enriched material and hence it consists of re-accreted material.


\section{Luminous star clusters in VV124} 
\label{clus}

\begin{figure*}
   \centering
   \includegraphics[width=\textwidth]{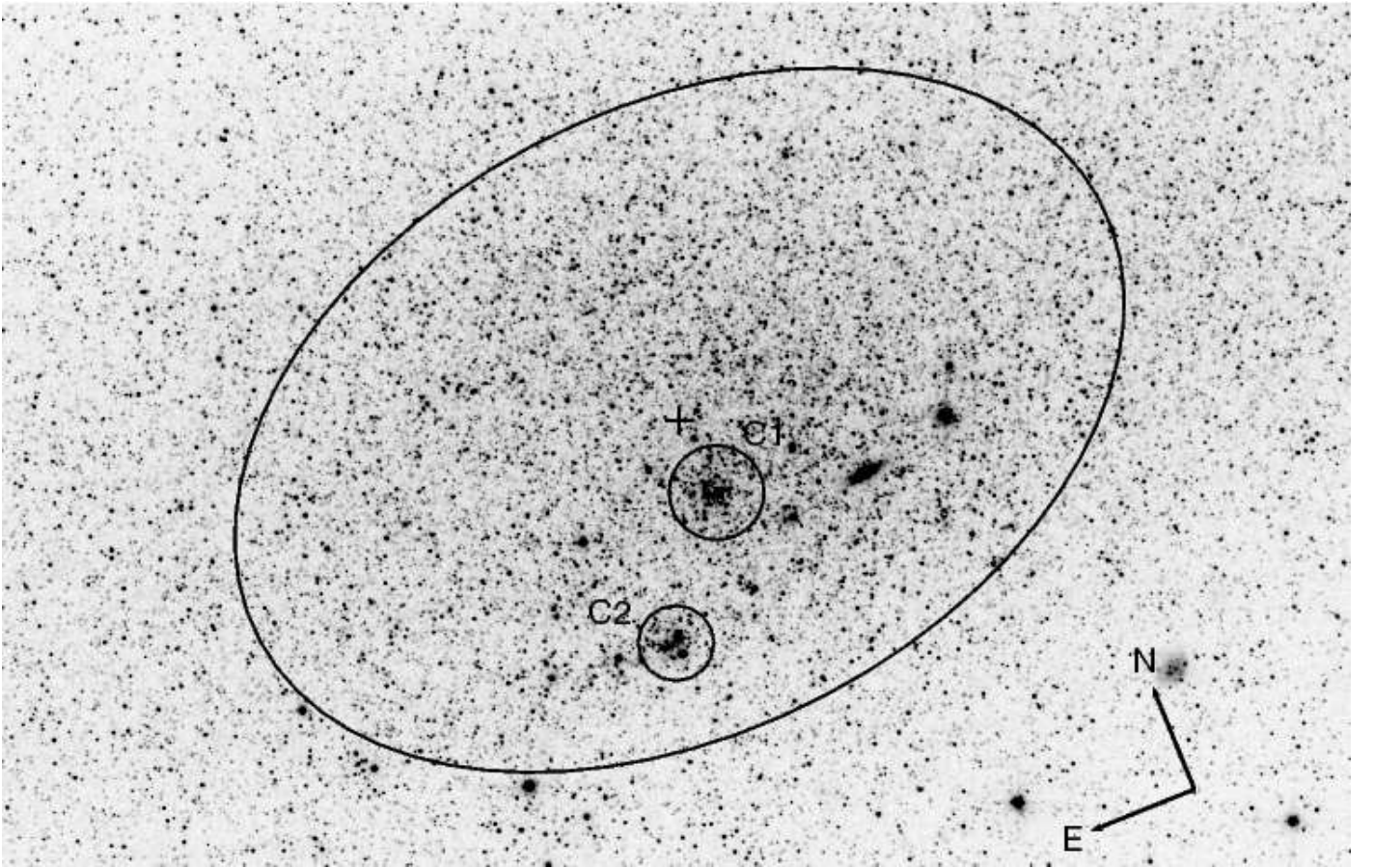}
     \caption{The two candidate clusters (C1 and C2) are indicated on this image of the center of VV124 (from the drizzled F606W image). The circle around C1 has a radius of $2.5\arcsec$, the one around C2 a radius of $2.0\arcsec$. The ellipse has the same ellipticity and orientation of the VV124 isophotes and semi-major axis $a=25\arcsec$. The center of the galaxy is indicated by a cross.}
        \label{clusima}
    \end{figure*}

The zoomed view of the galaxy center shown in Fig.~\ref{clusima} shows that the candidate star cluster identified in Pap-I (Cluster 1, C1 hereafter) is indeed largely resolved into stars in the ACS images, hence it is likely a genuine star cluster (see below). On the other hand, the two bright sources to the west of C1 are a background galaxy and a foreground star, respectively. There are several small groups of young stars in this regions that may be possibly classified as open clusters or associations. However, the only other group of size/luminosity comparable to C1 is the one indicated as Cluster~2 (C2) in the figure.  C2  appears to coincide with the position of the \HII\ region identified by \citet[][by means of a long slit optical spectrum]{k08}. With the available data it is not possible to establish whether the \HII\ region is associated to the cluster or it is just a chance superposition.

In Sect.~5.1 and 5.2, below, we will study in deeper detail these two small stellar systems.  The insight into the physical properties of the clusters is greatly hampered by stellar crowding which is maximized with respect to the surrounding field. Indeed, to reach $F814W\sim26.5$ in the very small areas enclosing the clusters, we had to relax the selection on the crowding parameter, accepting stars up to ${\tt crowd} <1.0$. In addition, the contamination by field stars provides additional problems to the interpretation of the CMDs. For these reasons, the best we can hope to obtain from the analysis of the clusters CMD are age estimates with uncertainties of order 20\%-50\%, also due to the uncertainty in the associated metallicity and the systematics associated with uncertainties in the theoretical models 
\citep{galla}. However such large age uncertainties translates into uncertainties of 10\%-20\% in the mass estimates obtained by converting total luminosity into mass \citep[see][and references therein]{vdb0,all}.

In the following analysis we will try to estimate ages by fitting the cluster CMD with BASTI isochrones, following \citet{vdb0}. As a reference value for the cluster metallicity we assume $Z=0.004$, i.e.\ the mean value estimated by J11 for the young population of VV124. Since the two clusters are clearly associated with this young component, this is by far the best guess on the metallicity of the clusters. While we will present only solutions with $Z=0.004$, we have attempted to fit the observed CMDs also with isochrones with metallicity ranging from $Z=0.001$ to $Z=0.02$ and we found that the $Z=0.004$ models always provide the best overall fit.
Estimates obtained with the \citet{leo} isochrones, for comparison, gives age values larger by $\sim$30\% that those obtained with BASTI isochrones (hence slightly larger mass estimates), keeping fixed any other parameter.
Finally, the good match in color between the MS of the models and of the clusters shows that 
the reddening value adopted for the whole galaxy is appropriate also for the clusters, indicating they are not affected by additional extinction.

\begin{table*}
  \begin{center}
  \caption{ $\frac{N_B}{N_R}$ ratios for the candidate star clusters and their control fields}
  \label{Tab_NB}
  \begin{tabular}{lcccccc}
    \hline
Cluster Name & Cluster & CF$_E$ & CF$_N$ & CF$_W$ & Min. Excess$^c$& Max. Excess$^d$\\
\hline
C1$^{a}$       &0.84$\pm$0.16 &0.11$\pm$0.04 &0.13$\pm$0.04 &0.28$\pm$ 0.06& $3.3\sigma$ & $4.4\sigma$ \\
C2$^{b}$       &0.70$\pm$0.16 &0.22$\pm$0.06 &0.13$\pm$0.04 &0.29$\pm$ 0.08& $2.3\sigma$ & $3.4\sigma$ \\
\hline
\end{tabular}
\tablefoot{$^a$ $N_B$ is the number of stars having $F606W-F814W\le 0.2$ and $24.0<F814W<26.0$, while $N_R$ is the number of stars having $F606W-F814W\ge 0.4$ in the same magnitude range.\\
$^b$ The definitions of $N_B$ and $N_R$ are the same as for C1, except for the magnitude range that  is $22.0<F814W<26.0$, in this case.\\
$^c$ Difference between the $\frac{N_B}{N_R}$ in the cluster and in the CF showing the lowest value of this parameter, in units of $\sigma$.\\
$^d$ Difference between the $\frac{N_B}{N_R}$ in the cluster and in the CF showing the highest value of this parameter, in units of $\sigma$.} 
\end{center}
\end{table*}

\subsection{Cluster~1}

\begin{figure}
   \centering
   \includegraphics[width=\columnwidth]{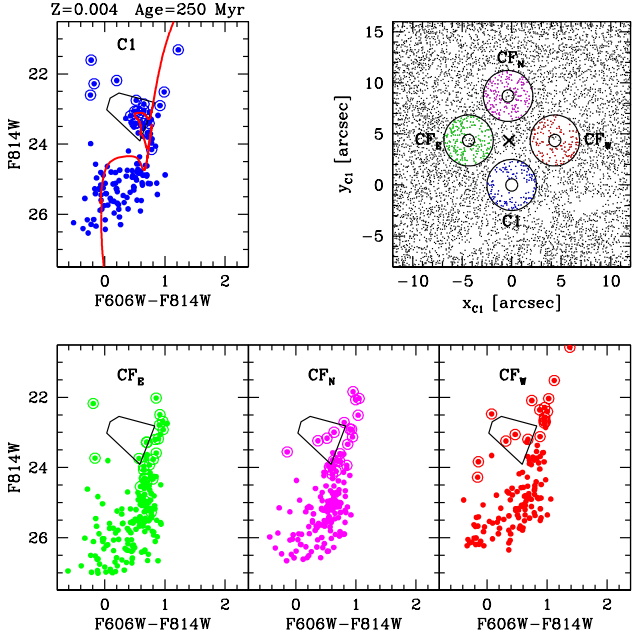}
     \caption{Comparison of the CMDs of the circular annuli around Cluster~1 and three nearby Control Fields indicated in the upper right panel. In this map the $\times$ sign indicates the center of the galaxy. The plotted points correspond to sources having ${\tt crowd}<1.0$. In the CMDs, sources with ${\tt crowd}<0.3$ are marked by an empty circle concentric to the filled one. An isochrone providing a reasonable fit to the observed CMD is also over-plotted to the C1 CMD (thick red line). A selection box for the BL population is superimposed to each CMD.}
        \label{cluster}
    \end{figure}

In Fig.~\ref{cluster} we compare the CMD in a small area (inner radius=$0.6\arcsec$, outer radius=$2.5\arcsec$) centered on C1 with those from identical areas centered on positions symmetric with respect to the center of VV124 and which are used as control fields (CF, see the upper right panel of the figure). It is quite clear that, even if the C1 region has a larger area devoid of measured stars because of the higher crowding associated to the cluster center, the C1 CMD show an excess of MS stars fainter than $F814W\simeq 24.5$ and, in particular, an {\em obvious} excess of BL stars, with respect to all the considered CFs. 
This is clearly observed even if the analysis is limited to stars with ${\tt crowd}<0.3$.
In particular in the C1 CMD there are 18 stars lying in the superposed BL selection box, while
3, 5, and 4 are found in the CF$_E$, CF$_N$, and CF$_W$ CMDs, respectively. The associated excess of BL stars in C1 with respect to the three CFs is at the 3.3$\sigma$, 2.7$\sigma$, and 3.0$\sigma$ level. 
If we limit to the sample of stars with ${\tt crowd}<0.3$, we find 11 stars in the BL box of the C1 CMD, while only 3 stars are found in the CMD of each CF.

In Table~\ref{Tab_NB} the ratio of stars bluer than $F606W-F814W=0.2$ ($N_B$) to the stars redder than $F606W-F814W=0.4$ ($N_R$) in the range $24.0<F814W<26.0$ is shown to be significantly larger (at a $>3\sigma$ level) in C1 than in any considered CF. Given the regular, compact  and roundish appearance of C1 and the obvious excess of young stars found in the C1 area, we conclude that C1 is a genuine star cluster. It is worth noting that the BL and the MS fainter than $F814W\sim 24.0$ are the only features of the C1 CMD that can be unequivocally attributed to the cluster, as they are significantly over-abundant in the C1 field than in the CFs. On the other hand, one can be tempted to interpret the five stars of the C1 CMD lying in the region $F606W-F814W<0.2$ and $F814W<24.0$ as very young MS stars associated to the cluster. However this interpretation has no statistical basis, since in the same region there are 3, 1, and 3 stars in CF$_E$,CF$_N$, and CF$_W$, respectively. Hence this handful of bright and blue stars in the C1 CMD is fully compatible with the expectations for the field population. Moreover, we did not find an isochrone that simultaneously matches these very blue stars and the BL population, that is unequivocally associated with C1.

In the upper left panel of Fig.~\ref{cluster} we show also the comparison of the observed CMD of C1 with the isochrone corresponding to our preferred solution.  In the age range covered by the young population of VV124, the luminosity of the Blue Loop is a very sensitive age indicator \citep{dohm}: the BL of C1 is well matched by the adopted isochrone, having an age of 250~Myr. Such a young age is in agreement with the integrated spectra of C1 we presented in Pap-I. Following \citet{vdb0}, we associate an uncertainty of $\pm50$~Myr to our age estimate. All age estimates we obtained by adopting isochrones of different metallicity (in the range Z=0.001 -0.02) lie within this error-bar.

We obtained estimates of the integrated V and I magnitudes from aperture photometry within a circular radius $r=3.0\arcsec$, corresponding to 15.7~pc\footnote{Integrated magnitudes have been transformed from F606W,F814W to V,I using the transformations by \citet{sirianni}} . These, as well as other relevant cluster parameters, are listed in Tab.~\ref{Tab_C1}. While the cluster may be slightly more extended than the adopted radius, this choice should be a reasonable compromise between including most of the cluster light and avoiding strong contamination from the dense field population. The sky level for aperture photometry was estimated in a nearby region, to subtract also the contribution from field stars of VV124. From aperture photometry on concentric annuli, and assuming $r=3.0\arcsec$ as the limiting radius of the cluster, we also obtained a rough estimate of the half-light radius. $r_h\simeq 1.5\arcsec$, corresponding to $\simeq 9.5$~pc, has been consistently obtained from both the F606W and the F814W images. This is within the range of $r_h$ values measured by \citet{barmby} in M31 clusters of similar age and mass (see below). 

The total V luminosity is $L_V=6.5\pm1.0\times 10^4~L_{\sun}$.
Assuming ${\frac{M}{L_V}}=0.188$, from a Simple Stellar Population \citep[SSP][]{rfp} model having $Z=0.004$, age = 250~Myr, we obtain $M_{*}=1.2\pm 0.2\times 10^4~M_{\sun}$\footnote{All the BASTI SPSS models adopt a \citet{kroupa} initial mass function (IMF).}.
This implies that C1 is significantly brighter than Galactic open clusters of similar age and it is  more massive than Galactic open clusters of any age \citep{open,vdb0}, except for very few (and much younger) young massive clusters discussed in \citet{figer} and \citet{mary}. On the other hand, it appears very similar in age, mass and size to the young massive clusters found in relatively large numbers in many star forming galaxies of any mass \citep[see][for references and discussion]{lr99,all,barmby,vs06,lr11}.
In the cluster classification scheme adopted by \citet{lr00} and \citet{billett}
C1 clearly belongs to the class of {\em regular} star clusters \citep[$M_V<-8.5$, see][]{annibal}.

\begin{table}
  \begin{center}
  \caption{Observed and derived parameters of cluster C1}\label{Tab_C1}
  \begin{tabular}{lcr}
    \hline
    Parameter & value & notes\\
\hline
$\alpha_0$     & 09:16:03.14                    & J2000$^*$  \\
$\delta_0$     & +52:50:27.0                    & J2000$^*$ \\
age            & $250\pm 50$ Myr               & $^a$   \\
$[M/H]$        & $-0.7$                         & $^a$ \\
R$_p$          & $4.3\arcsec$                   & $^b$ \\
$r_h$          & $\sim 1.5\arcsec$                  &  $^c$ \\
$V_{tot}$      & $18.4 \pm 0.1$                & $^d$  \\
$I_{tot}$      & $18.1 \pm 0.1$                & $^d$  \\
$M_V$          & $-7.2 \pm 0.16$                &  \\
$L_V$          & $6.5\pm1.0\times 10^4~L_{\sun}$    & total V luminosity \\
$M_{stars}$    & $1.2\pm 0.2\times 10^4~M_{\sun}$      & stellar mass$^e$  \\
\hline
\end{tabular}
\tablefoot{$^*$ In the astrometric system embedded into the F606W drizzled fits image.
$^a$ From the isochrone fit of Fig.~\ref{cluster} (See Sect.~5).
$^b$ Projected distance from the center of VV124.
$^c$ From aperture photometry on concentric annuli, assuming a limiting radius $r_l=3.0\arcsec$.
$^d$ From aperture photometry with radius $r=3.0\arcsec$.
$^e$ Assuming $(M/L)_V=0.188$ from the BASTI database, for a model having Age=250~Myr, Z=0.004 and a Kroupa IMF.
} 
\end{center}
\end{table}

\subsection{Cluster~2}

C2 appears more irregular in shape with respect to C1, being clearly dominated by a few bright stars. This suggests a younger age than that of C1.
We performed the same kind of analysis as done for the latter cluster. The $\frac{N_B}{N_R}$ ratio indicates an overdensity of young stars in C2 with respect to all the CFs, albeit at a lower level of significance than in the case of C1 (Tab~\ref{Tab_NB}). We conclude that also C2 is likely a genuine cluster. 

It is very hard to age-date this cluster. If the \HII\ region is indeed physically associated with C2, the age is constrained to be $\la 10$~Myr, and the three stars brighter than $F814W=21.0$ are not cluster BL stars. This is clearly possible since, even if two of them are redder than $F606W-F814W=0.0$ and  do not have direct counterpart in the CMD of the CFs, this may be merely due to fluctuations inherent with such small number of stars.

On the other hand, if these stars are interpreted as genuine cluster BL stars, the age should be larger than the above value and the coincidence with the \HII\ region must be considered as due to chance superposition. Unfortunately, the accuracy of the positioning of the \HII\ region is not sufficient to check
the degree of spatial coincidence with C2 on scales smaller than $1\arcsec-2\arcsec$ \citep[see Fig.~2 in][]{k08}. Considering the three brightest stars as cluster members, the best fit is obtained with an isochrone of age 30$\pm 10$~Myr. Given the above discussion, we take this value as an upper limit to the age of C2.

\begin{table}
  \begin{center}
  \caption{Observed and derived parameters of cluster C2}\label{Tab_C2}
  \begin{tabular}{lcr}
    \hline
    Parameter & value & notes\\
\hline
$\alpha_0$     & 09:16:03.69                    & J2000$^*$  \\
$\delta_0$     & +52:50:20.4                    & J2000$^*$ \\
age            & $\le 30\pm10$ Myr                 & $^a$   \\
$[M/H]$        & $-0.7$                         & $^a$ \\
R$_p$          & $11.9\arcsec$                   & $^b$ \\
$r_h$          & $\sim 0.6\arcsec$                  &  $^c$ \\
$V_{tot}$      & $18.6 \pm 0.1$                & $^d$  \\
$I_{tot}$      & $18.6 \pm 0.1$                & $^d$  \\
$M_V$          & $-7.0 \pm 0.16$                &  \\
$L_V$          & $5.4\pm 0.8\times 10^4~L_{\sun}$    & total V luminosity \\
$M_{stars}$    & $\le 3.3\pm 0.5\times 10^3~M_{\sun}$      & stellar mass$^e$  \\
\hline
\end{tabular}
\tablefoot{$^*$ In the astrometric system embedded into the F606W drizzled fits image.
$^a$ From the isochrone fit of Fig.~\ref{cluster} (See Sect.~5).
$^b$ Projected distance from the center of VV124.
$^c$ From aperture photometry on concentric annuli, assuming a limiting radius $r_l=2.0\arcsec$.
$^d$ From aperture photometry with radius $r=2.0\arcsec$.
$^e$ Assuming $(M/L)_V=0.061$ from the BASTI database, for a model having Age=30~Myr, Z=0.004 and a Kroupa IMF.
} 
\end{center}
\end{table}

The total V luminosity is $L_V=5.4\pm 0.8\times 10^4~L_{\sun}$, the half-light radius is $r_h\simeq 0.6\arcsec$, corresponding to $\sim 4$~pc (all the estimates as been obtained in the same way as for C1, see above).
Assuming ${\frac{M}{L_V}}=0.061$, from a SSP model having Z=0.004, age=30~Myr from the BASTI set , we obtain $M_{*}\le 3.3\pm 0.7\times 10^3~M_{\sun}$.
This is in the high-mass side of the Galactic open clusters distribution, but not out of the range covered by those stellar systems. Also C2 should be included among {\em regular} star clusters \citep{lr00,billett}.

   \begin{figure}
   \centering
   \includegraphics[width=\columnwidth]{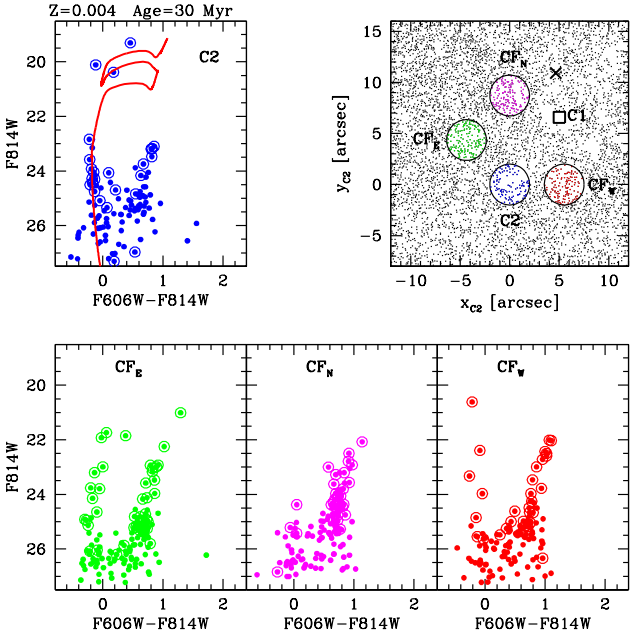}
     \caption{The same as Fig.~\ref{cluster} for Cluster~2. In this case the CF$_W$ field is not symmetric with respect to the other CFs, to avoid inclusion of  a significant portion of C1, whose position is also indicated in the map.}
        \label{cluster2}
    \end{figure}


\section{Summary and Conclusions}
\label{conc}

We have presented a new reduction and analysis of archival HST-ACS images of
the isolated dwarf galaxy VV124. A detailed study of the SFH in the innermost regions of the galaxy, from the same data,  has been already presented by another group (J11). Here we have focused our attention on a few specific issues that have been raised in Pap-I and were not addressed in the J11 paper. 

\begin{itemize} 

\item We provide a clear detection of an extended Horizontal Branch population. It has been shown that  
HB stars have a spatial distribution which is less concentrated than the RC population. This strongly supports the idea that very old and metal poor stars become more and more dominant at larger distances from the center of VV124. 

\item The near coincidence of the asymmetric sheet of young stars near the center of VV124 and the density peak of the (also asymmetric)  associated \HI cloud strongly support the idea that the complex kinematics and velocity gradients observed in the gas in Pap-I are most likely associated with the dynamics of the star formation episode. It is interesting to note that some of the isolated dwarfs in the simulations by \citet{S11} lose all their \HI at very early times and are able to re-accrete some of the expelled gas at later times. This scenario would be fully consistent with the observed properties of the \HI in VV124 and would provide a natural explanation for the long period of quiescence between the early phase of star formation and the very recent sprout of activity that originated the sparse population of young MS stars currently observed (see J11).

\item We confirm that the candidate cluster identified in Pap-I (C1) is a genuine young massive star cluster \citep[see][and references therein]{all}. It has an age of $\sim 250$~Myr and a total mass of 
$\simeq 1.2\times10^4~M_{\sun}$. The cases of such low-mass early-type dwarf galaxies with massive clusters are quite rare in the Local Group (M98). We identified another likely cluster (C2), significantly younger (age$\le 30$~Myr) and less massive ($M\le 3.3\times10^3~M_{\sun}$) than C1. 
 
\end{itemize}

\begin{acknowledgements}
We warmly thank Andy Dolphin for his precious assistance in the data analysis.
We are also grateful to M. Tosi, A. Sollima and M. Cignoni for a critical reading of a draft version of the manuscript.
M.B. and S.P. acknowledges the partial financial
support to this research by INAF through the PRIN-INAF 2009 grant CRA 1.06.12.10 (PI: R. Gratton).
Partial financial support for this study was also provided by ASI through contracts COFIS ASI-INAF I/016/07/0 and ASI-INAF I/009/10/0. 

\end{acknowledgements}

\bibliographystyle{apj}



\end{document}